# Digital spiral object identification using random light

Zhe Yang[1,2], Omar S. Magaña-Loaiza[2,†], Mohammad Mirhosseini[2], Yiyu Zhou[2], Boshen Gao[2], Lu Gao[2,5], Seyed Mohammad Hashemi Rafsanjani[2], Guilu Long[1,4,†] and Robert W. Boyd[2,3]

Photons that are entangled or correlated in orbital angular momentum have been extensively used for remote sensing, object identification and imaging. It has recently been demonstrated that intensity fluctuations give rise to the formation of correlations in the orbital angular momentum components and angular positions of random light. Here, we demonstrate that the spatial signatures and phase information of an object with rotational symmetries can be identified using classical orbital angular momentum correlations in random light. The Fourier components imprinted in the digital spiral spectrum of the object, as measured through intensity correlations, unveil its spatial and phase information. Sharing similarities with conventional compressive sensing protocols that exploit sparsity to reduce the number of measurements required to reconstruct a signal, our technique allows sensing of an object with fewer measurements than other schemes that use pixel-by-pixel imaging. One remarkable advantage of our technique is that it does not require the preparation of fragile quantum states of light and operates at both low- and high-light levels. In addition, our technique is robust against environmental noise, a fundamental feature of any realistic scheme for remote sensing.

1 State Key Laboratory of Low-dimensional Quantum Physics and Department of Physics, Tsinghua University, Beijing 100084, China

2 The Institute of Optics, University of Rochester, Rochester, New York, 14627 USA

3 Department of Physics, University of Ottawa, Ottawa, ON K1N 6N5, Canada

4 Tsinghua National Laboratory for Information Science and Technology, Beijing 100084, China

5 School of Science, China University of Geosciences, Beijing 100083, China

†Correspondence: Omar S. Magaña-Loaiza, E-mail: omar.maganaloaiza@rochester.edu

†Correspondence: Guilu Long, E-mail: gllong@mail.tsinghua.edu.cn

# INTRODUCTION

The orbital angular momentum (OAM) of light has attracted considerable attention in recent years. As identified by Allen *et al.* in 1992, a beam of light with an azimuthal phase dependence of the form $e^{-il\phi}$ carries OAM, where $l$ is the mode index, which specifies the amount of OAM, and $\phi$ is the azimuthal angle[1]. This interesting property of light has been explored in different contexts. For example, fundamental tests of high-dimensional entangled systems have been performed through the OAM basis[2], the infinite OAM bases have been used to implement paradoxes in quantum mechanics[3], and relativistic effects have been explored in the azimuthal degree of freedom[4,5]. In the applied context, the OAM of light has been used to encode information[6-10], manipulate microscopic-particles[11-14], perform optical metrology[15,16], and perform remote sensing and imaging[4,17-25].

It has been suggested that the discrete OAM spectrum (or spiral spectrum) can be used for imaging, a technique known as digital spiral imaging[17]. In addition, quantum OAM correlations[26] have been used to enhance the image contrast of phase objects[18]. Furthermore, object identification has been performed by means of quantum correlated OAM states[19,20]. Similarly, quantum correlations have been incorporated into digital spiral imaging to retrieve information of phase objects[21]. Moreover, field correlations in vectorial beams have been utilized for kinematic sensing[22].

It has been recently demonstrated that random fluctuations of light give rise to the formation of intensity correlations in the OAM components and angular positions of pseudothermal light[27]. It has also been shown that these classical correlations are manifested through interference structures that resemble those observed with entangled photons. These results suggest that OAM intensity correlations in random optical fields, such as those found in natural light, could be used to develop optical technologies with the similar functionality as those that employ entangled photons.

In this work, we exploit the OAM correlations of random light to demonstrate object identification; in this approach, the object is identified through its discrete OAM spectrum (or spiral spectrum). We also demonstrate that we can use the same types of correlations to retrieve the phase information of an object. Despite the fact that intensity correlations in the OAM degree of freedom are not perfect, as for the case of entangled photons, it is possible to perform object identification at any light levels, an important advantage over the quantum protocols that employ fragile entangled states of light.

# MATERIALS AND METHODS

**Theoretical analysis**

The OAM spectrum of a random light field $E(r,\phi)$ can be experimentally measured by projecting it onto a series of OAM modes $l$. The amplitude for this projection is given by

$$a_l = \int r dr d\phi \, E(r,\phi) \frac{e^{-il\phi}}{\sqrt{2\pi}}. \tag{1}$$

The angular coherence properties of a field of light are described by the first-order correlation function

$G^{(1)}(1,1) = \langle a_1^* a_1 \rangle = \langle I_1 \rangle$, where the symbol $\langle ... \rangle$ indicates the ensemble average. Similarly, the second-order correlation function that describes intensity correlations in the OAM domain is defined as $G^{(2)}(1_1, 1_2) = \langle I_{1_1}^* I_{1_2} \rangle = \langle a_{1_1}^* a_{1_1} a_{1_2}^* a_{1_2} \rangle$. For a thermal beam of light, $G^{(2)}(1_1, 1_2)$ is given by[27,28]

$$G^{(2)}(1_1, 1_2) = G^{(1)}(1_1, 1_1) G^{(1)}(1_2, 1_2) + \left| G^{(1)}(1_1, 1_2) \right|^2 \quad (2)$$

The first term in Eq. (2) represents a product of intensities between two OAM modes. This first term is constant and represents a background that causes the intensity correlations not to be perfect, whereas the second term, $\left| G^{(1)}(1_1, 1_2) \right|^2$, is typically approximated by a delta function that describes point-to-point OAM correlations.

In our scheme for object identification, one of the two beams illuminates the object described by the transmission function $A(r, \phi)$. In this case, the second term of Eq. (2), which is defined as $\Delta G^{(2)}(1_1, 1_2) \equiv \left| G^{(1)}(1_1, 1_2) \right|^2$, can be expressed (see the Supplementary Materials) as

$$\Delta G^{(2)}(1_1, 1_2) = \left| \int r dr d\phi \, \overline{\left| E(r, \phi) \right|^2} A(r, \phi) \frac{e^{i\Delta l \phi}}{2\pi} \right|^2, \quad (3)$$

where $\Delta l = 1_1 - 1_2$. Interestingly, the object $A(r, \phi)$ encodes its Fourier components into the second-order correlation function. This signature is used to recover its spatial or phase information. When the object is not present, $A(r, \phi) = 1$, and this term takes the form of a $\delta$-function.

**Experimental Setup**

Fig. 1a shows the experimental setup we use for digital spiral object identification. A 532-nm diode laser illuminates a digital micro-mirror device (DMD), which is used to generate a random field of light[29,30]. A 4*f*-optical system consisting of two lenses and a spatial filter is employed to isolate the first order of the beam diffracted by the DMD. The intensity distribution of the generated beam is shown in Fig. 1b.

The random light field is divided into "test" and "reference" arms after passing through a beam splitter. The light beam in the test arm interacts with an amplitude or phase object, which is displayed onto a spatial light modulator (SLM), as shown in Fig. 1c. Each light beam is then projected onto a forked hologram to measure an OAM component of the random field of light[19-21,31]. The first diffraction order of the structured beam is filtered by an aperture and then is coupled into a single mode fiber (SMF) and detected by an avalanche photodiode (APD). Two APDs and a coincidence count module are utilized to measure OAM correlations between the two arms. The total accumulation time of each measurement is set to 5 s in our experiment.

**RESULTS AND DISCUSSION**

**Amplitude Object Identification**

As shown in Figs. 2a and b, we use objects with four- and six-fold rotational symmetries. Each object is encoded onto the SLM located in the test arm.

A series of OAM projections are performed in each arm to construct a two-dimensional matrix with the normalized second-order correlation function; see Figs. 2c and d. The OAM number in the test and reference arms are denoted by $l_t$ and $l_r$, respectively. The normalized second-order OAM correlation function is calculated by $g^{(2)}(l_t, l_r) = \langle I_{l_t} I_{l_r} \rangle / \langle I_{l_t} \rangle \langle I_{l_r} \rangle$, where $\langle I_{l_t} I_{l_r} \rangle$ is proportional to the coincidence count rate. Each element in the matrix is obtained by averaging over 50 realizations of the experiment, and the error bars are obtained by calculating the standard deviation.

As shown in Figs. 2c and d, an amplitude object with $N$-fold rotational symmetry imprints its Fourier components into the second-order OAM correlation matrix. The correlation signal is high along the diagonal elements of the matrix, where $\Delta l = l_t - l_r = \pm N$ due to the symmetry of the amplitude object. In our case, these signatures can be observed when $\Delta l = \pm 4$ for the object with four-fold rotational symmetry and when $\Delta l = \pm 6$ for the object with six-fold rotational symmetry. Consequently, it is evident that one can use the OAM correlation matrix to identify the two objects. Furthermore, note that this technique requires a small number of measurements compared to traditional imaging schemes that rely on pixel-by-pixel raster scanning.

In Figs. 2e and f, we plot the transverse sections, defined by $g^{(2)}(l_t, l_r = 0)$, for the correlation matrices in Figs. 2c and d, respectively. For simple and symmetric objects, a single line in the correlation matrix can provide adequate information about the object. However, the measurement of the total OAM correlation matrix is required for complicated objects that lack rotational symmetry[19,32].

**Phase Object Identification**

We showed above that our technique is capable of identifying amplitude objects with rotational symmetry. Next, we demonstrate that our technique can also be used to identify phase objects. As a specific example, we use phase objects consisting of non-integer vortices described as $e^{-iM\phi}$, where $M$ indicates a non-integer winding number[33-37]. The phase profile of a vortex with $M=-2/3$ is shown in Fig. 3a; the azimuthal phase for a non-integer vortex of this form ranges from $-2\pi/3$ to $2\pi/3$. The forked hologram that we encode onto the SLM is shown in Fig. 3b. The two-dimensional normalized second-order OAM correlation matrix is shown in Fig. 3c, and its middle row is plotted in Fig. 3d. In this case, the presence of the phase object induces a broader spectrum in the correlation matrix.

As shown in Fig. 4, we also test the performance of our technique with different phase objects characterized by the non-integer winding numbers $M=-1/2$, $M=-5/2$, $M=-2/3$ and $M=-8/3$. The performance of our technique can be characterized through the Floor function. This simple function is used to denote the largest previous integer of $M$ and can be defined as $u = \lfloor M \rfloor$, and $v$ is the non-integer part given by $v = M - u$. The theoretical and experimental results show that the central peak of the correlation signal is determined by $u$ and the profile is determined by $v$. A simple comparison between Fig. 4a and Fig. 4b shows that the two figures have the same profile, but the central peak is located at different positions. This difference is because the parameter $v$ is equal to 1/2 for both cases, whereas the parameter $u$ is different; this parameter is equal to -1 and -3 for Fig. 4a and Fig. 4b, respectively. We can compare the results shown in Fig. 4c and Fig. 4d; in this case, the parameter $v$ is equal to 1/3, whereas the parameter $u$ is equal to -1 and -3 for Fig. 4c and Fig. 4d, respectively. In this case, the two figures have the same profile, but the peak is located at different positions.

In our experiment, we used phase objects consisting of non-integer vortices. However, this technique can be applied to the identification of other phase objects, such as those discussed in Refs [38-40]. These schemes require coherent sources of light or entangled photons.

The use of random light and intensity correlations in our technique shares similarities with other techniques, such as thermal ghost imaging[41]. However, our technique extracts the fingerprints that characterize objects with rotational symmetries, leading to a reduction in the number of measurements required in conventional techniques for imaging. Another interesting aspect of our technique is that the second-order interference effects are less sensitive to the coherence properties of the source. In fact, this is one of the advantages of the Hanbury Brown and Twiss interferometer compared to the Michelson interferometer[42]. In addition, it has been demonstrated that imaging schemes based on second-order correlations are robust against turbulence[43].

## CONCLUSIONS

We experimentally demonstrated digital spiral object identification for an amplitude and a phase object using second-order OAM correlations with random light. In our technique, the object imprints its Fourier components onto the digital spiral spectrum; by measuring intensity correlations in the OAM degree of freedom, we can retrieve spatial and phase information for different masks. Compared to conventional pixel-by-pixel imaging, this technique only requires a small fraction of the number of measurements to identify an object; this peculiarity makes our technique sparse sensitive, similar to other techniques that rely on compressive sensing. In addition, our technique does not rely on fragile quantum states of light and can operate at low- and high-light levels. Finally, our technique is robust against environmental noise and has potential applications in remote sensing and imaging.

## AUTHOR CONTRIBUTIONS

O.S.M.L. conceived the idea. The experiment was designed by Z.Y, O.S.M.L., M.M., G. L. and R.W.B. The theoretical description of our work was developed by Z.Y., B. G. and S.M.H.R. The experiment was performed by Z.Y., Y. Z., L. G., O.S.M.L. and M.M. The data were analyzed by Z. Y., with help from O.S.M.L. The project was supervised by G.L. and R.W.B. All authors contributed to the discussion of the results and to the writing of the manuscript.

## ACKNOWLEDGMENTS

We gratefully acknowledge Jiapeng Zhao for valuable discussions. We also thank Jiying Jia for revising the manuscript. Zhe Yang is thankful for the financial support from the program of the China Scholarship Council (No.201506210145). Lu Gao acknowledges the support from the National Natural Science Foundation of China, No.11504337. GLL acknowledges the partial support from the Natural Science Foundation of China under Grant Nos. 11175094 and 91221205 and the National Basic Research Program of China under Grant No 2015CB921002. GLL is a member of the Center of Atomic and Molecular Nanosciences, Tsinghua University.

**REFERENCES**

1. Allen L, Beijersbergen MW, Spreeuw RJC, Woerdman JP. Orbital angular momentum of light and the transformation of Laguerre-Gaussian laser modes. *Phys Rev A* 1992; **45**: 8185-8189.
2. Molina-Terriza G, Torres JP, Torner L. Twisted photons. *Nat Phys* 2007; **3**: 305-310.
3. Potoček V, Miatto FM, Mirhosseini M, Magaña-Loaiza OS, Liapis AC *et al*. Quantum Hilbert hotel. *Phys Rev Lett* 2015; **115**: 160505.
4. Bialynicki-Birula I, Bialynicka-Birula Z. Rotational frequency shift. *Phys Rev Lett* 1997; **78**: 2539-2542.
5. Lavery MP, Speirits FC, Barnett SM, Padgett MJ. Detection of a spinning object using light's orbital angular momentum. *Science* 2013; **341**: 537-540.
6. Wang J, Yang JY, Fazal IM, Ahmed N, Yan Y *et al*. Terabit free-space data transmission employing orbital angular momentum multiplexing. *Nat Photonics* 2012; **6**: 488-496.
7. Willner AE, Wang J, Huang H. A different angle on light communications. *Science* 2012; **337**: 655-656.
8. Bozinovic N, Yue Y, Ren YX, Tur M, Kristensen P *et al*. Terabit-scale orbital angular momentum mode division multiplexing in fibers. *Science* 2013; **340**: 1545-1548.
9. Mirhosseini M, Magaña-Loaiza OS, O'Sullivan MN, Rodenburg B, Malik M *et al*. High-dimensional quantum cryptography with twisted light. *New J Phys* 2015; **17**: 033033.
10. Zhang CX, Guo BH, Cheng GM, Guo JJ, Fan RH. Spin-orbit hybrid entanglement quantum key distribution scheme. *Sci China Phys Mech Astron* 2014; **57**: 2043-2048.
11. Friese MEJ, Nieminen TA, Heckenberg NR, Rubinsztein-Dunlop H. Optical alignment and spinning of laser-trapped microscopic particles. *Nature* 1998; **394**: 348-350.
12. Paterson L, MacDonald MP, Arlt J, Sibbett W, Bryant PE *et al*. Controlled rotation of optically trapped microscopic particles. *Science* 2001; **292**: 912-914.
13. O'Neil AT, MacVicarI, Allen L, Padgett MJ. Intrinsic and extrinsic nature of the orbital angular momentum of a light beam. *Phys Rev Lett* 2002; **88**: 053601.
14. Grier DG. A revolution in optical manipulation. *Nature* 2003; **424**: 810-816.
15. D'Ambrosio V, Spagnolo N, Del Re L, Slussarenko S, Li Y *et al*. Photonic polarization gears for ultra-sensitive angular measurements. *Nat Commmon* 2013; **4**: 2432.
16. Magaña-Loaiza OS, Mirhosseini M, Rodenburg B, Boyd RW. Amplification of angular rotations using weak measurements. *Phys Rev Lett* 2014; **112**: 200401.
17. Torner L, Torres JP, Carrasco S. Digital spiral imaging. *Opt Express* 2005; **13**: 873-881.
18. Jack B, Leach J, Romero J, Franke-Arnold S, Ritsch-Marte M *et al*. Holographic ghost imaging and the violation of a Bell inequality. *Phys Rev Lett* 2009; **103**: 083602.
19. Uribe-Patarroyo N, Fraine A, Simon DS, Minaeva O, Sergienko AV. Object identification using correlated orbital angular momentum states. *Phys Rev Lett* 2013; **110**: 043601.
20. Simon DS, Sergienko AV. Two-photon spiral imaging with correlated orbital angular momentum states. *Phys Rev A* 2012; **85**: 043825.
21. Chen LX, Lei JJ, Romero J. Quantum digital spiral imaging. *Light Sci Appl* 2014; **3**: e153.

22  Berg-Johansen S, Töppel F, Stiller B, Banzer P, Ornigotti M *et al*. Classically entangled optical beams for high-speed kinematic sensing. *Optica* 2015; **2**: 864-868.

23  Cvijetic N, Milione G, Ip E, Wang T. Detecting lateral motion using light's orbital angular momentum. *Sci Rep* 2014; **5**: 15422.

24  Milione G, Cvijetic N, Wang T. System and method for remote object sensing. U.S. patent 2016. No. 20,160,146,937.

25  Cvijetic N, Milione G, Ip E, Wang T. Method and apparatus for remote sensing using optical orbital angular momentum (OAM)-based spectroscopy for detecting lateral motion of a remote object. U.S. Patent Application 2015. No. 14/975,252.

26  Leach J, Jack B, Romero J, Jha AK, Yao AM *et al*. Quantum correlations in optical angle–orbital angular momentum variables. *Science* 2010; **329**: 662-665.

27  Magaña-Loaiza OS, Mirhosseini M, Cross RM, Rafsanjani SMH, Boyd RW. Hanbury brown and twiss interferometry with twisted light. *Sci Adv* 2016; **2**: e1501143.

28  Xiong J, Cao DZ, Huang F, Li HG, Sun XJ *et al*. Experimental observation of classical subwavelength interference with a pseudothermal light source. *Phys Rev Lett* 2005; **94**: 173601.

29  Rodenburg B, Mirhosseini M, Magaña-Loaiza OS, Boyd RW. Experimental generation of an optical field with arbitrary spatial coherence properties. *J Opt Soc Am B* 2014; **31**: A51-A55.

30  Mirhosseini M, Magaña-Loaiza OS, Chen CC, Rodenburg B, Malik M *et al*. Rapid generation of light beams carrying orbital angular momentum. *Opt Express* 2013; **21**: 30196-30203.

31  Shapira A, Naor L, Arie A. Nonlinear optical holograms for spatial and spectral shaping of light waves. *Sci Bull* 2015; **60**: 1403-1415.

32  Fitzpatrick CA, Simon DS, Sergienko AV. High-capacity imaging and rotationally insensitive object identification with correlated orbital angular momentum states. *Int J Quantum Inform* 2014; **12**: 1560013.

33  Leach J, Yao E, Padgett MJ. Observation of the vortex structure of a non-integer vortex beam. *New J Phys* 2004; **6**: 71.

34  Oemrawsingh SSR, Ma X, Voigt D, Aiello A, Eliel ER *et al*. Experimental demonstration of fractional orbital angular momentum entanglement of two photons. *Phys Rev Lett* 2005; **95**: 240501.

35  Götte JB, Franke-Arnold S, Zambrini R, Barnett SM. Quantum formulation of fractional orbital angular momentum. *J Mod Opt* 2007; **54**: 1723-1738.

36  Dennis MR, O'Holleran K, Padgett MJ. Singular optics: optical vortices and polarization singularities. *Prog Opt* 2009; **53**: 293-363.

37  O'Dwyer DP, Phelan CF, Rakovich YP, Eastham PR, Lunney JG *et al*. Generation of continuously tunable fractional optical orbital angular momentum using internal conical diffraction. *Opt Express* 2010; **18**: 16480-16485.

38  Yao E, Franke-Arnold S, Courtial J, Barnett S, Padgett MJ. Fourier relationship between angular position and optical orbital angular momentum. *Opt Express* 2006; **14**: 9071-9076.

39  Jha AK, Jack B, Yao E, Leach J, Boyd RW *et al*. Fourier relationship between the angle and angular momentum of entangled photons. *Phys Rev A* 2008; **78**: 043810.

40  Jack B, Padgett MJ, Franke-Arnold S. Angular diffraction. *New J Phys* 2008; **10**: 103013.

41  Pittman TB, Shih YH, Strekalov DV, Sergienko AV. Optical imaging by means of two-photon quantum entanglement. *Phys Rev A* 1995; **52**: R3429-R3432.

42  Born M, Wolf E. *Principles of Optics*. 7th edn. New York: Cambridge University Press; 1999. p986.

43  Dixon PB, Howland GA, Chan KWC, O'Sullivan-Hale C, Rodenburg B *et al*. Quantum ghost imaging through turbulence. *Phys Rev A*, 2011; **83**: 051803.

# List of Figure Captions

**Figure 1**. (**a**) Experimental setup. A DMD is illuminated by a 532-nm laser beam. The first diffraction order of the structured beam is isolated by a 4*f*-optical system comprised of two lenses and a spatial filter in the focal plane (figure not to scale). A series of random patterns are displayed on the DMD at a frequency of 1.4 kHz to produce a random field of light. The generated beam is divided by a beam splitter to produce a test beam that interacts with the object and a reference beam. An SLM in each arm is used to measure the OAM components in the random beam of light. (**b**) Image of the spatial intensity distribution of the random beam of light. (**c**) The amplitude or phase object is encoded into the SLM in the test arm.

**Figure 2**. Digital spiral identification for amplitude objects with four- and six-fold rotational symmetries. An object with four-fold rotational symmetry with $\alpha = \pi/6$ and $\beta = \pi/4$ is depicted in (**a**). A similar object with $\alpha = \pi/8$ and $\beta = \pi/3$ is shown in (**b**). The corresponding second-order correlation matrices are shown in (**c**) and (**d**). The rows denoted by the dotted boxes in (**c**) and (**d**) are plotted in (**e**) and (**f**), respectively.

**Figure 3**. Digital spiral identification for a phase object. (**a**) The phase object consisting on a non-integer vortex with an OAM number of *M*=-2/3. (**b**) The corresponding forked hologram that we encode onto the SLM located in the test beam. (**c**) Experimental results for the second-order OAM correlation matrix. (**d**) A plot of the row denoted by the dotted box in (**c**).

**Figure 4**. Digital spiral identification for phase objects with different non-integer winding numbers: (a) *M*=-1/2, (b) *M*=-5/2, (c) *M*=-2/3, and (d) *M*=-8/3.

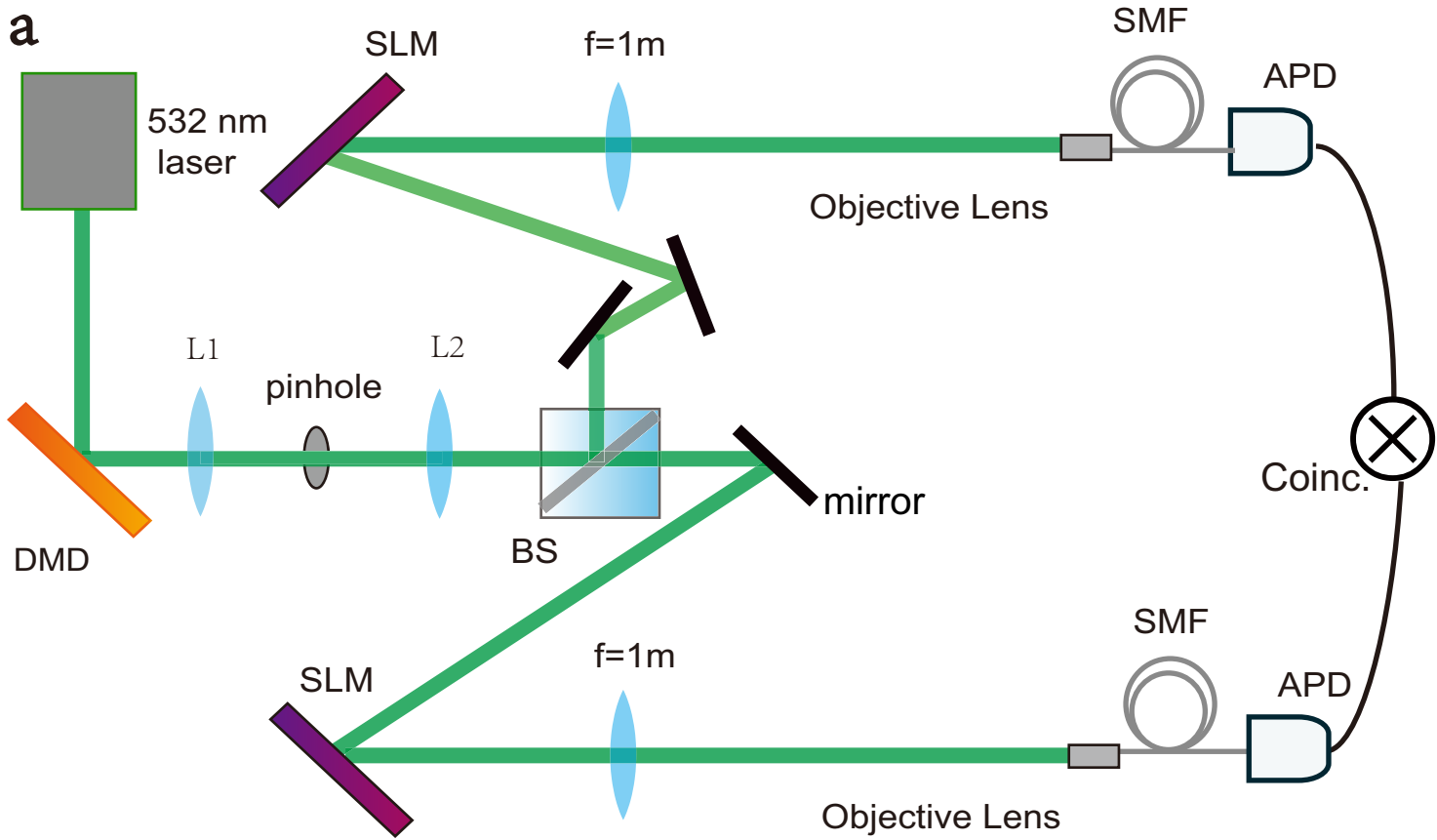

a

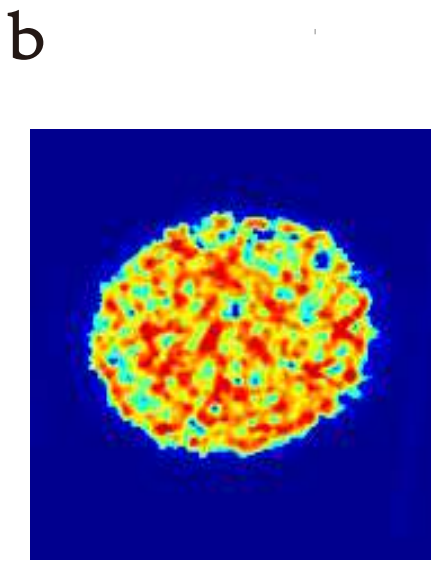

b

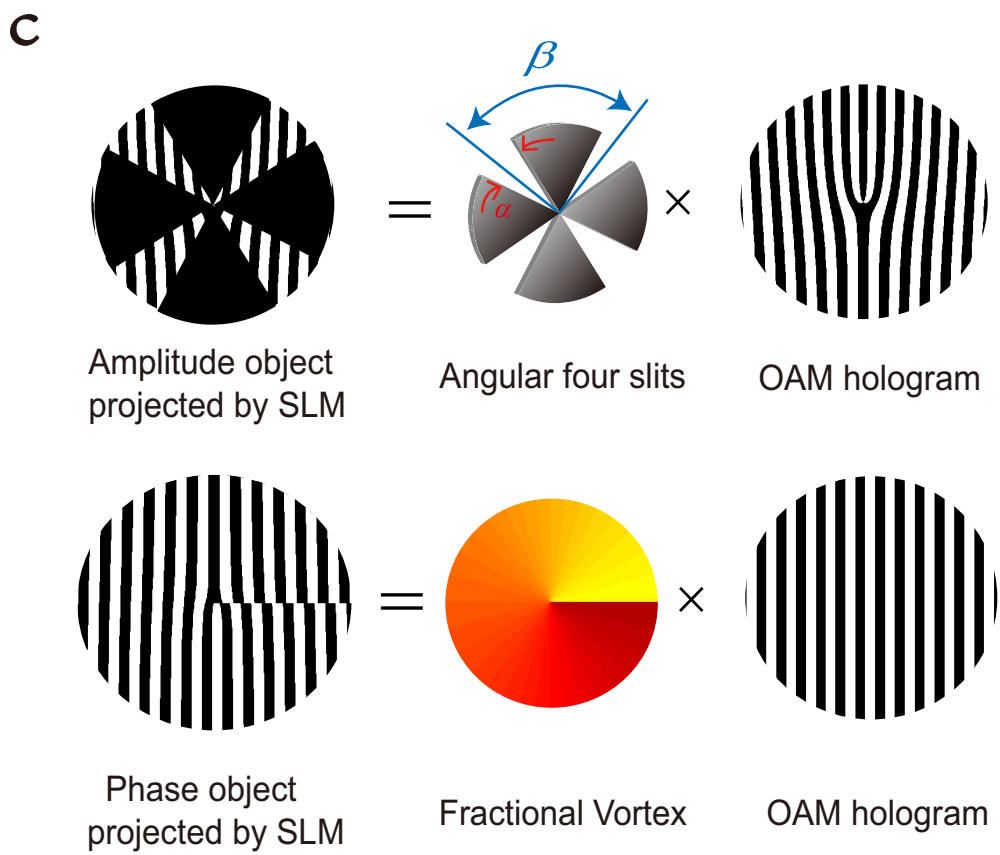

c

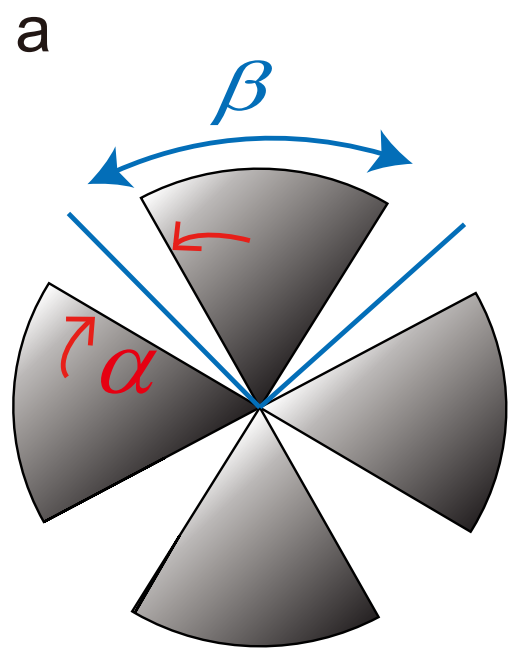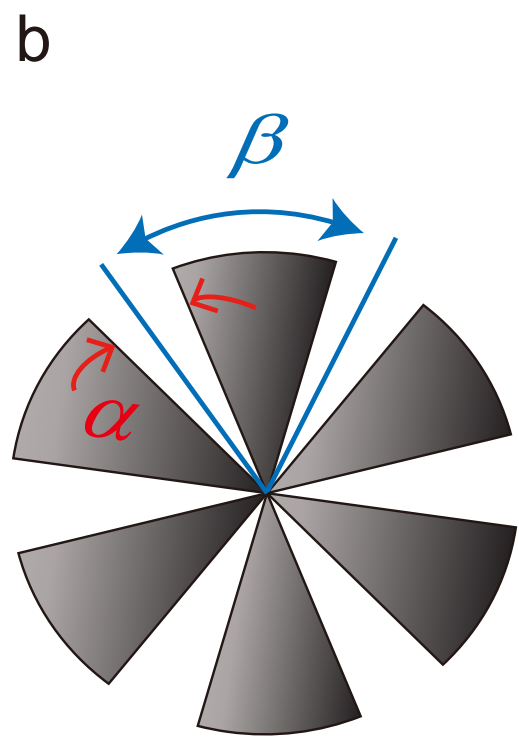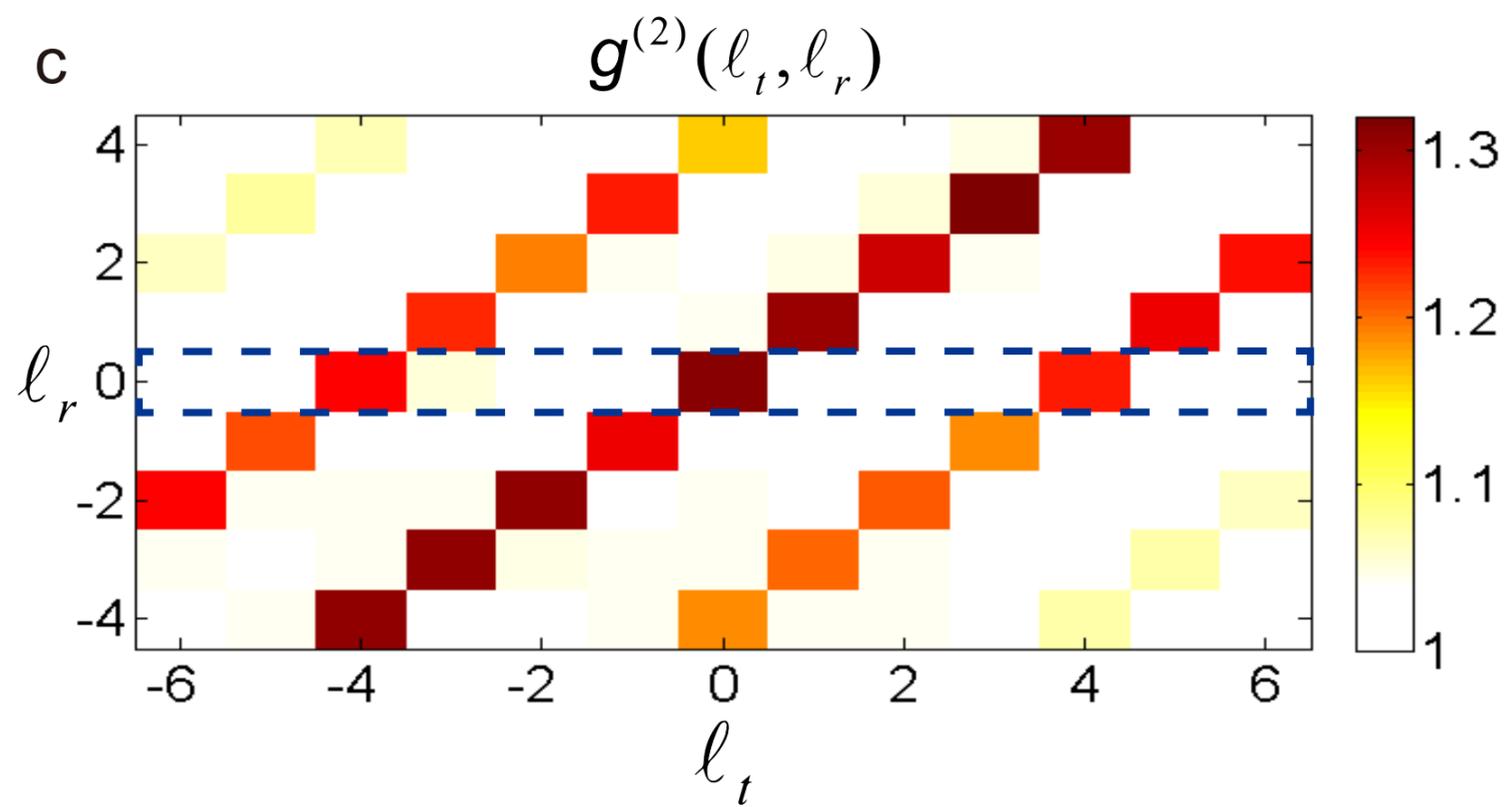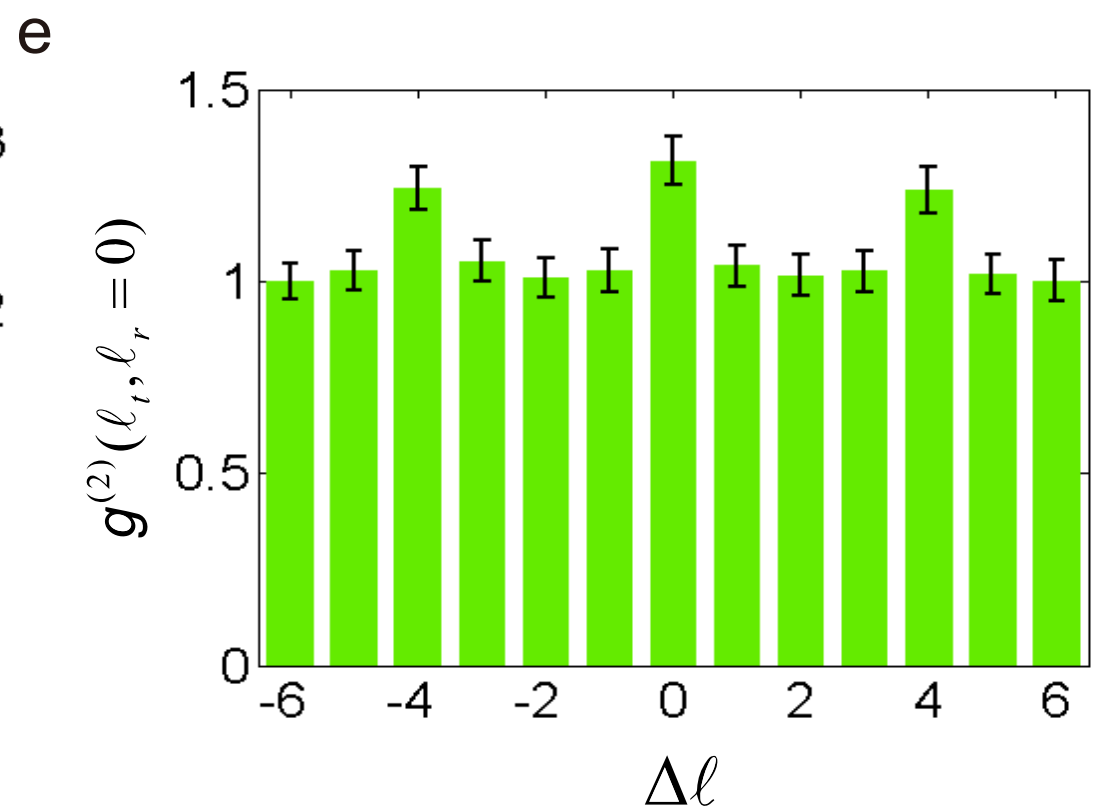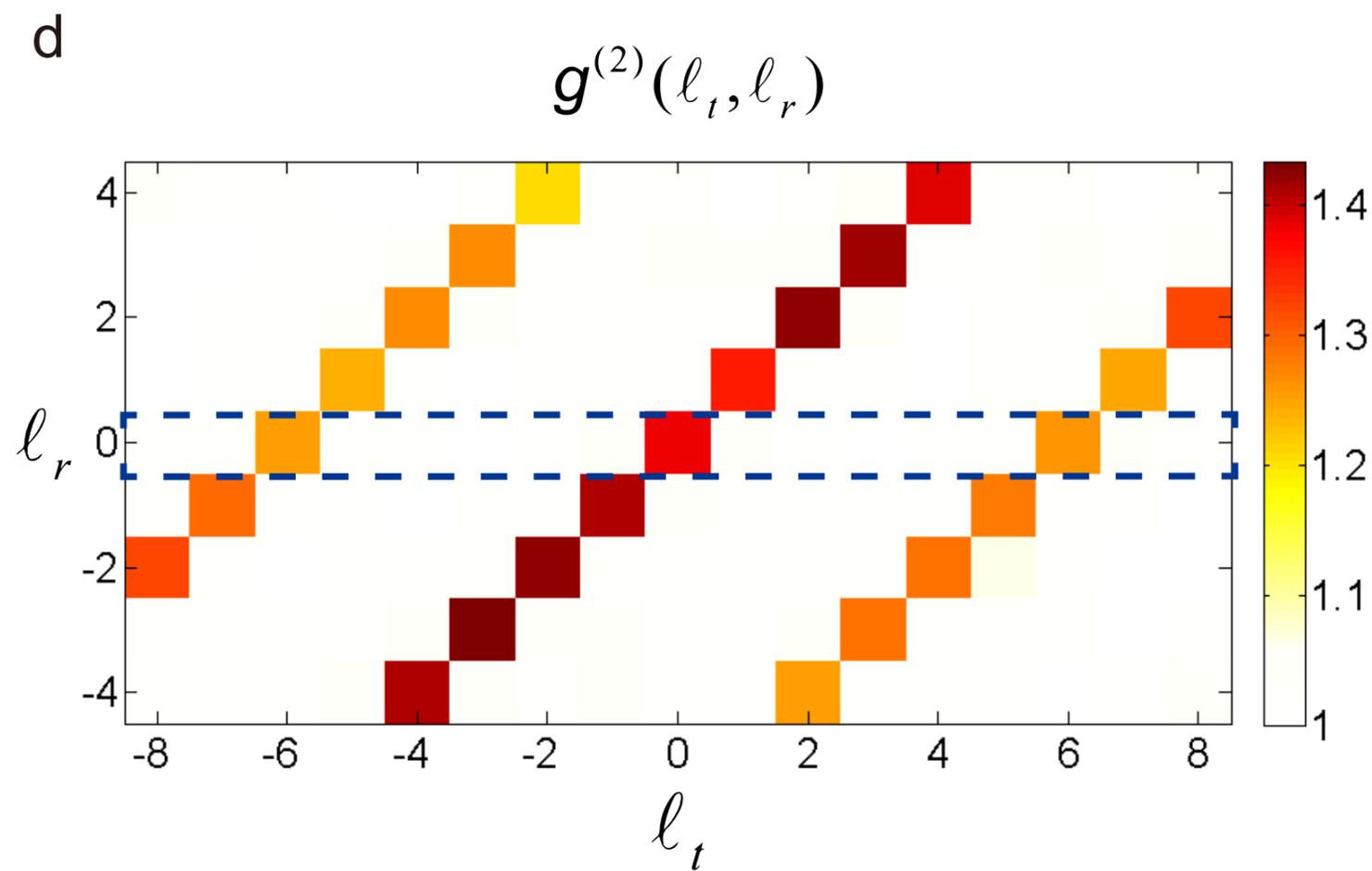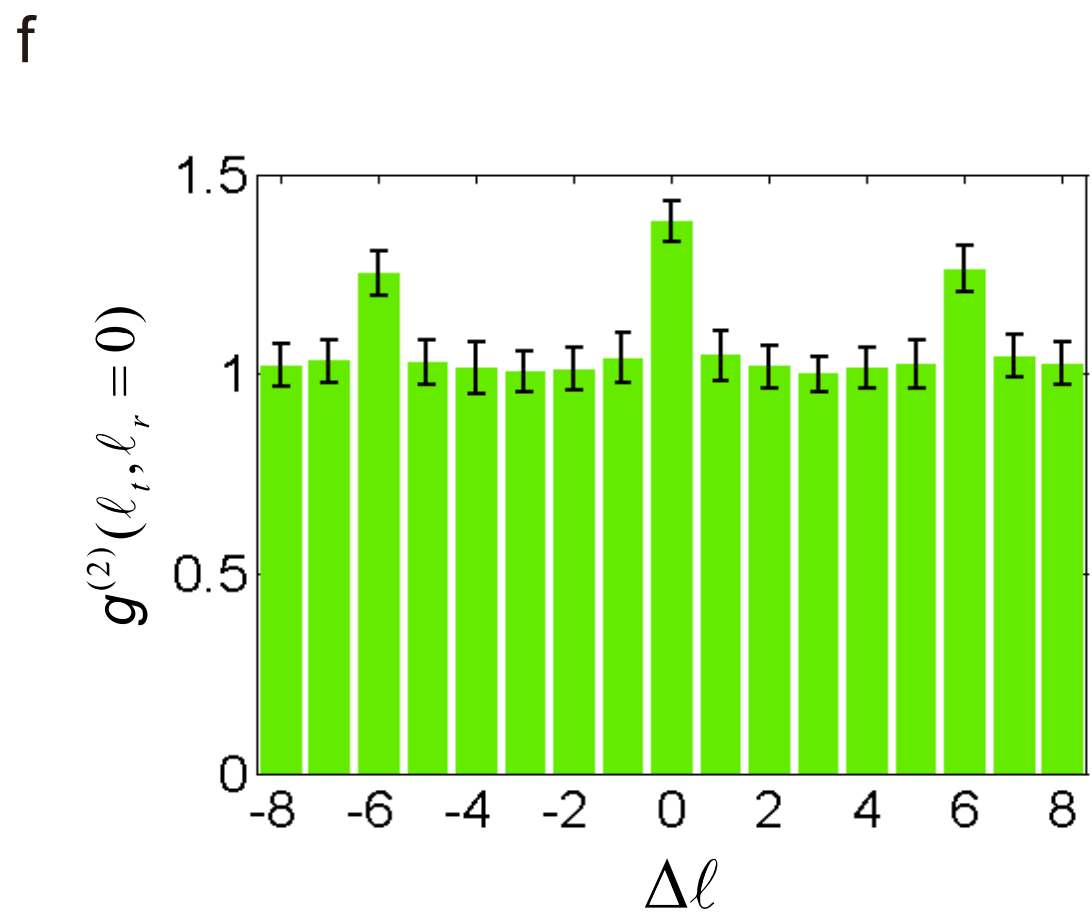

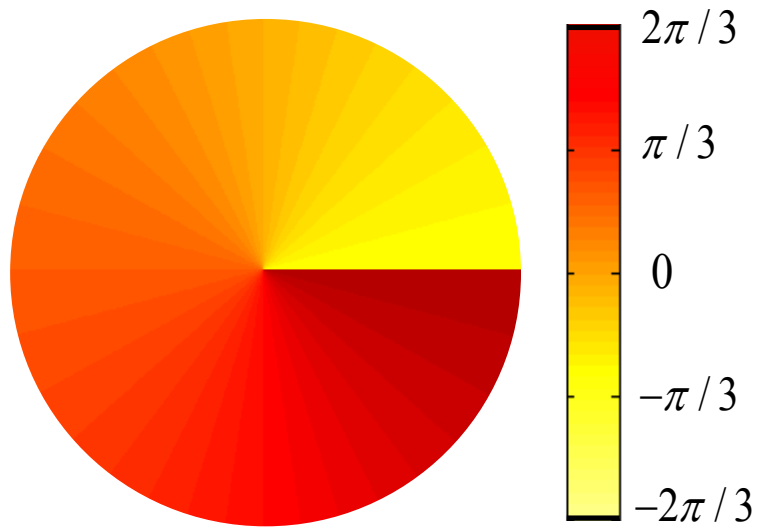
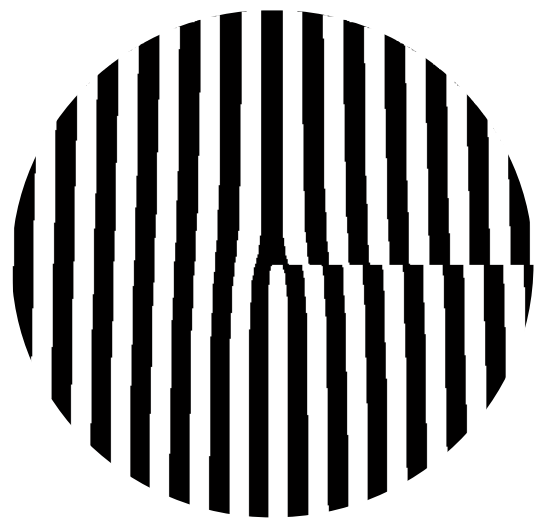
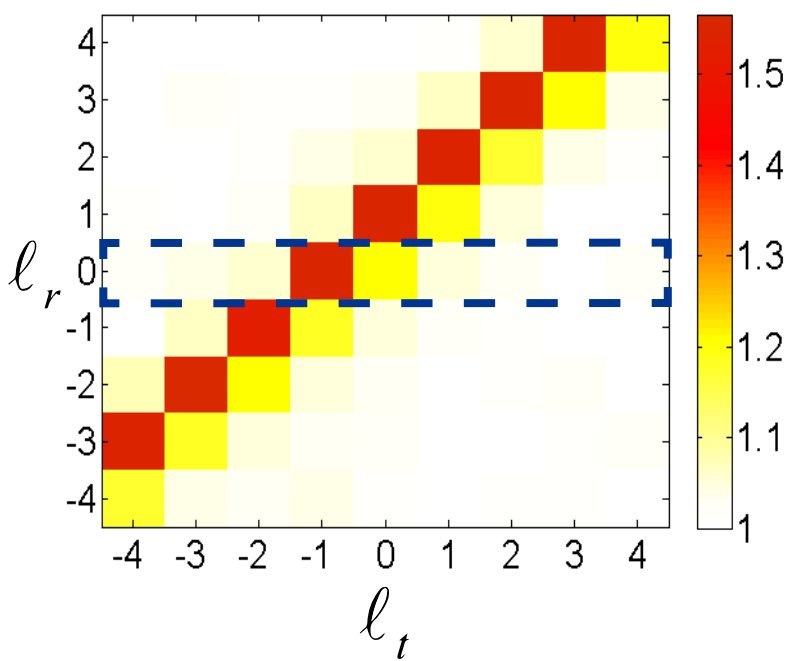
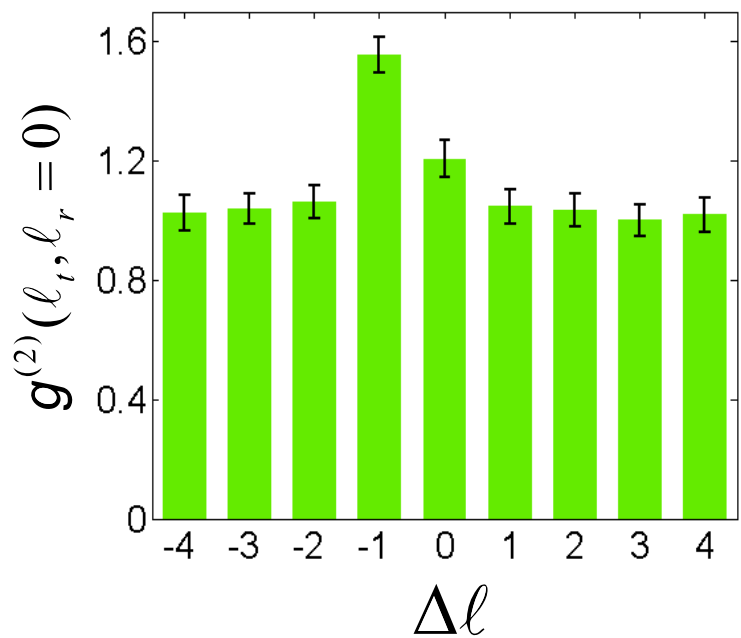

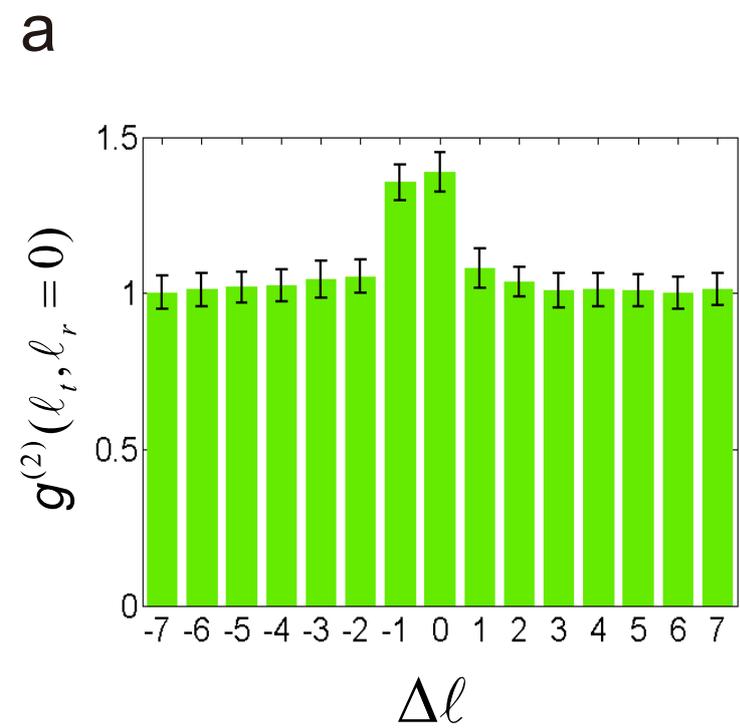

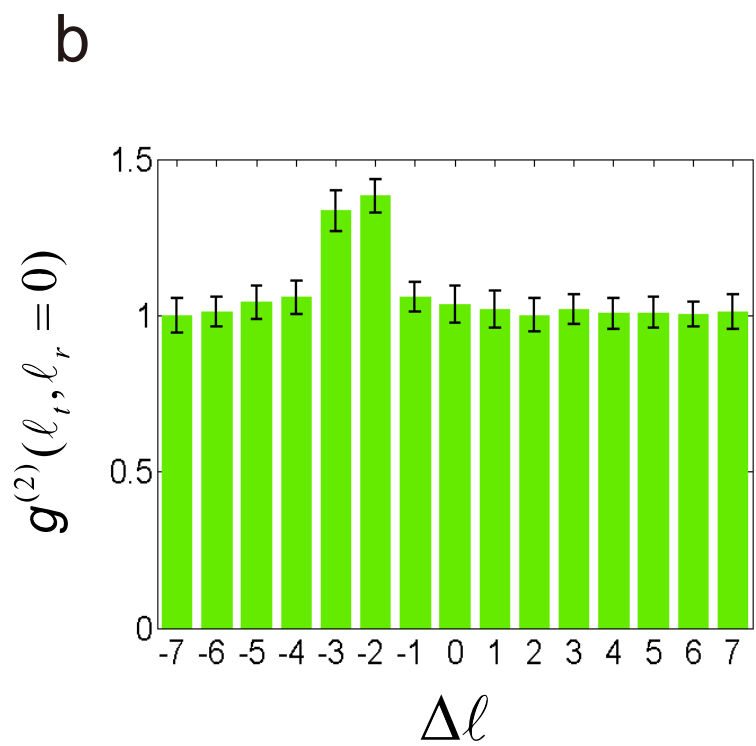

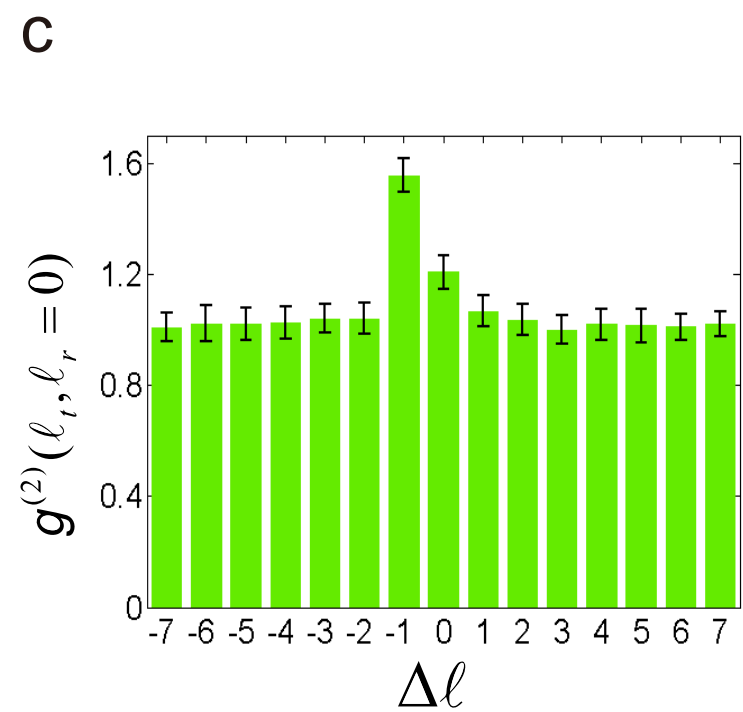

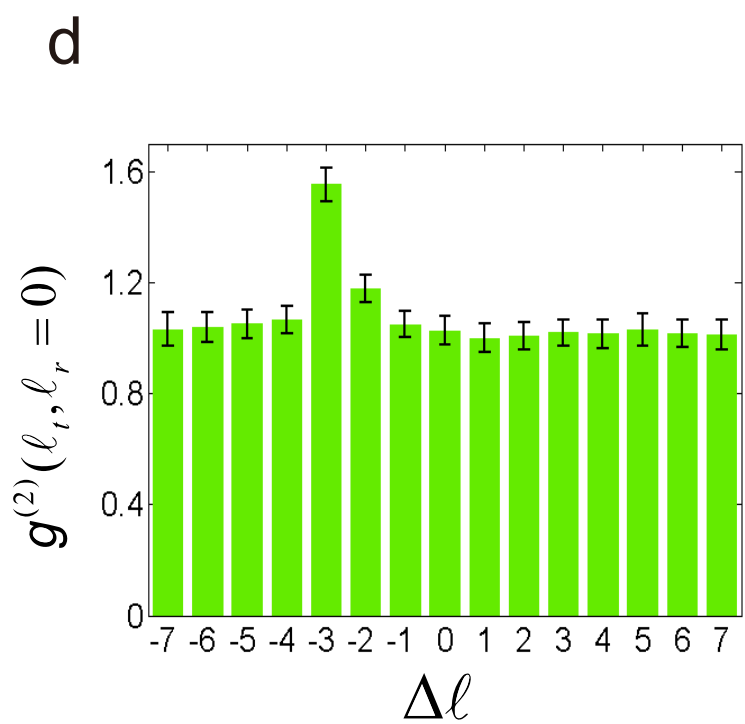

# SUPPLEMENTAL METERIAL:

## Digital spiral object identification using random light

Zhe Yang[1,2], Omar S. Magaña-Loaiza[2,†], Mohammad Mirhosseini[2], Yiyu Zhou[2], Boshen Gao[2], Lu Gao[2,5], Seyed Mohammad Hashemi Rafsanjani[2], Guilu Long[1,4,†] and Robert W. Boyd[2,3]

### I. Second-order OAM correlations between two beams

We describe a thermal light field in the polar coordinate by $E(r,\phi)$. Experimentally, the thermal light field can be projected onto a series of OAM modes l to measure the corresponding OAM spectrum. The field amplitude of this measurement is given by

$$a_1 = \int r dr d\phi \, E(r,\phi) \frac{e^{-il\phi}}{\sqrt{2\pi}}. \tag{1}$$

The first-order OAM correlation function $G^{(1)}(1,1) = \langle a_1^* a_1 \rangle = \langle I_1 \rangle$ describes the angular coherence properties of the light field.

If $l_1 = l_2$, $G^{(1)}(l,l)$ represents the intensity of the light field projected onto the OAM mode l

$$I_1 = a_1^* a_1 = \int r_1 dr_1 d\phi_1 \, E^*(r_1,\phi_1) \frac{e^{il\phi_1}}{\sqrt{2\pi}} \times \int r_2 dr_2 d\phi_2 \, E(r_2,\phi_2) \frac{e^{-il\phi_2}}{\sqrt{2\pi}}. \tag{2}$$

[1] State Key Laboratory of Low-dimensional Quantum Physics and Department of Physics, Tsinghua University, Beijing 100084, China

2 The Institute of Optics, University of Rochester, Rochester, New York, 14627 USA

3 Department of Physics, University of Ottawa, Ottawa, ON K1N 6N5, Canada

4 Tsinghua National Laboratory for Information Science and Technology, Beijing 100084, China

5 School of Science, China University of Geosciences, Beijing 100083, China

†Correspondence: Omar S. Magaña-Loaiza, E-mail: omar.maganaloaiza@rochester.edu

†Correspondence: Guilu Long, E-mail: gllong@mail.tsinghua.edu.cn

The ensemble average spectral intensity can be described as

$$\langle I_1 \rangle = \int r_1 dr_1 d\phi_1 r_2 dr_2 d\phi_2 \frac{e^{il(\phi_1 - \phi_2)}}{2\pi} \langle E^*(r_1,\phi_1) E(r_2,\phi_2) \rangle. \tag{3}$$

If the transverse coherence length of the random light field is small enough, one get the relation

$$\langle E^*(r_1,\phi_1) E(r_2,\phi_2) \rangle = \overline{|E(r_1,\phi_1)|^2} \frac{\delta(r_1 - r_2)\delta(\phi_1 - \phi_2)}{r_1}. \tag{4}$$

The ensemble average spectral intensity can be expressed as

$$\langle I_1 \rangle = \frac{1}{2\pi} \int r dr d\phi \, \overline{|E(r,\phi)|^2}, \tag{5}$$

which is independent on winding number $l$ of OAM.

If an object is placed in the light beam, the light field takes the form of $E(r,\phi)A(r,\phi)$ where $A(r,\phi)$ is the transmission function of the object. The thermal light field passing through an arbitrary amplitude and phase object can be expressed in terms of angular harmonics that $E(r,\phi)A(r,\phi) = \sum_l a_l e^{il\phi} / \sqrt{2\pi}$. The coefficients $a_l$ can be calculated by $a_l = \int r dr d\phi \, E(r,\phi) A(r,\phi) e^{-il\phi} / \sqrt{2\pi}$. Therefore, the average spectral intensity of this beam reads as

$$\langle I_1 \rangle = \frac{1}{2\pi} \int r dr d\phi \, \overline{|E(r,\phi)|^2} |A(r,\phi)|^2. \tag{6}$$

Similar to Eq. (5), this term is also independent on $l$, consequently, one can not obtain any information of the object from the digital spiral spectrum of a single thermal beam of light.

The second-order OAM correlation between two beams is defined as $G^{(2)}(l_1, l_2) = I_{l_1} I_{l_2}$, and this takes the form of

$$I_{l_1} I_{l_2} = a_{l_1}^* a_{l_1} a_{l_2}^* a_{l_2} =$$

$$\int r_1 dr_1 d\phi_1 \, E^*(r_1,\phi_1) A^*(r_1,\phi_1) \frac{e^{il_1 \phi_1}}{\sqrt{2\pi}} \times \int r_2 dr_2 d\phi_2 \, E(r_2,\phi_2) A(r_2,\phi_2) \frac{e^{-il_1 \phi_2}}{\sqrt{2\pi}} \tag{7}$$

$$\times \int r_3 dr_3 d\phi_3 \, E^*(r_3,\phi_3) \frac{e^{il_2 \phi_3}}{\sqrt{2\pi}} \times \int r_4 dr_4 d\phi_4 E(r_4,\phi_4) \frac{e^{-il_2 \phi_4}}{\sqrt{2\pi}}.$$

Therefore, the ensemble average of the second-order OAM correlation reads as

$$\langle I_{l_1} I_{l_2}\rangle = \langle a^*_{l_1} a_{l_1} a^*_{l_2} a_{l_2}\rangle =$$

$$\int r_1 dr_1 d\phi_1 \int r_2 dr_2 d\phi_2 \int r_3 dr_3 d\phi_3 \int r_4 dr_4 d\phi_4 \ A^*(r_1,\phi_1) A(r_2,\phi_2) \qquad (8)$$

$$\frac{e^{il_1\phi_1}}{\sqrt{2\pi}} \frac{e^{-il_1\phi_2}}{\sqrt{2\pi}} \frac{e^{-il_2\phi_3}}{\sqrt{2\pi}} \frac{e^{il_2\phi_4}}{\sqrt{2\pi}} \langle E^*(r_1,\phi_1) E(r_2,\phi_2) E^*(r_3,\phi_3) E(r_4,\phi_4)\rangle.$$

For a thermal light field, the second-order correlation can be expressed in terms of first-order correlation functions

$$\langle E^*(r_1,\phi_1) E(r_2,\phi_2) E^*(r_3,\phi_3) E(r_4,\phi_4)\rangle$$
$$= \langle E^*(r_1,\phi_1) E(r_2,\phi_2)\rangle \langle E^*(r_3,\phi_3) E(r_4,\phi_4)\rangle + \langle E^*(r_1,\phi_1) E(r_4,\phi_4)\rangle \langle E(r_2,\phi_2) E^*(r_3,\phi_3)\rangle. \qquad (9)$$

Combining Eq. (8) and Eq. (9), one can obtain the second-order OAM correlation

$$G^{(2)}(l_1,l_2) = G^{(1)}(l_1,l_1) G^{(1)}(l_2,l_2) + |G^{(1)}(l_1,l_2)|^2, \qquad (10)$$

where the first term is the background which describes the ensemble average spectral intensities of two beams. The second term, which is given by $G^{(2)}(l_1,l_2) = G^{(1)}(l_1,l_1) G^{(1)}(l_2,l_2) + |G^{(1)}(l_1,l_2)|^2$, is the signal taking the form of

$$\Delta G^{(2)}(l_1,l_2) = \left| \int rdrd\phi \overline{|E(r,\phi)|^2} A(r,\phi) \frac{e^{i\Delta l \phi}}{2\pi} \right|^2, \qquad (11)$$

where $\Delta l = l_1 - l_2$.

## II. Identification for amplitude object

Now we discuss the identification for an amplitude object with *N*-fold rotational symmetry. The transmission function of the object with *N* periodic slits takes the form of

$$A(\phi) = \begin{cases} 1, & \text{for } n\beta, \ \phi < \alpha + n\beta \\ 0, & \text{else} \end{cases}, \qquad (12)$$

where *n=0,1,2,...N-1*, $\alpha$ is the width of one angular slit and $\beta = 2\pi/N$ is the spacing between the centers of two adjacent angular slits.

Substituting Eq. (12) into Eq. (11), the signal term is given by

$$G^{(2)}(l_1,l_2) \sim \langle I_{l_1}\rangle\langle I_{l_2}\rangle \left|\sum_{n=0}^{N-1} e^{i\frac{n\Delta l\beta}{2}}\right|^2 \frac{\alpha^2}{(2\pi)^2}\sin c^2(\frac{\Delta l\alpha}{2}). \tag{13}$$

The amplitude object with *N*-fold rotational symmetry imprints its particular signature to the OAM spectrum, from which we can be obtained the information of the object.

### III. Identification for phase object

In the following analysis, the non-integer vortex (fractional vortex) is treated as the phase object. The non-integer vortex can be described by $A(\phi)=e^{iM\phi}$, where *M* is not an integer and the corresponding second-order OAM correlation signal term is given by

$$\Delta G^{(2)}(1_1,1_2) = \left|\int rdrd\phi\, \overline{|E(r,\phi)|^2} e^{-iM\phi} \frac{e^{i\Delta l\phi}}{2\pi}\right|^2. \tag{14}$$

If the Floor function is used to denote the largest previous integer of *M*, i.e., $u=\lfloor M \rfloor$, and $v=M-u$ is the proper fractional part, one can arrive at

$$G^{(2)}(1_1,1_2) \sim \langle I_{1_1}\rangle\langle I_{1_2}\rangle \frac{2-2\cos(v\cdot 2\pi)}{(\Delta l-u-v)^2}. \tag{15}$$

This equation shows that the peak position of this term is determined by *u* and the spread distribution in the OAM spectrum is given by *v*.